\documentclass[prl,reprint,amsmath,amssymb,aps,longbibliography,superscriptaddress]{revtex4-2}

\usepackage{xcolor}
\usepackage{graphicx}
\usepackage{hyperref}
\usepackage{physics}
\usepackage{siunitx}
\usepackage{ulem}

\usepackage{dcolumn}
\usepackage{graphicx}
\usepackage{mdwlist}
\usepackage{subfigure}
\usepackage{booktabs}
\usepackage{amsmath}
\usepackage{extarrows}
 \usepackage{multirow}
\usepackage{dsfont}
\usepackage{amstext}
\usepackage{amsbsy}
\usepackage{bbm}
\usepackage{bm}
\usepackage{threeparttable}
\usepackage{booktabs}
\usepackage{bbding,pifont}
\usepackage{amsthm}
\usepackage{graphicx}
\usepackage{textcomp}
\usepackage{multirow}
\usepackage{pifont}
\usepackage{color}
\usepackage{sansmath}
\usepackage{makecell}
\usepackage{tabularx}

\hypersetup{
	colorlinks=true,
	urlcolor=purple,
	citecolor=blue,
	linkcolor=purple,
	breaklinks,
    }

\begin{document}

\title{Heralded Entanglement Transfer from Entangled Atomic Pair to Free Electrons}
\date{\today}

\author{Du Ran}
\email{randu11111@163.com}
\affiliation{School of Electronic Information Engineering, Yangtze Normal University, Chongqing 408100, China\\}
\affiliation{Key Laboratory of Optical Chip and Intelligent Optoelectronic Systems, Chongqing Municipal Education Commission, Yangtze Normal University,  Chongqing 408100, China\\}

\author{Reuven Ianconescu}
\affiliation{Department of Electrical Engineering Physical Electronics,Tel Aviv University, Ramat Aviv 69978, Israel\\}
\affiliation{Shenkar College of Engineering and Design 12, Anna Frank St., Ramat Gan, Israel\\}

\author{Shuai Liu}
\affiliation{School Of Physics And Electronic Engineering, Hubei University Of Arts And Science, Xiangyang, 441053, China\\}

\author{Ya-dong Li}
%\email{lyd1989@yznu.edu.cn}
\affiliation{School of Electronic Information Engineering, Yangtze Normal University, Chongqing 408100, China\\}

\author{Ji-Yuan Bai}
\affiliation{School of Electronic Information Engineering, Yangtze Normal University, Chongqing 408100, China\\}

\author{Ze-Long He}
\affiliation{School of Electronic Information Engineering, Yangtze Normal University, Chongqing 408100, China\\}
\affiliation{Key Laboratory of Optical Chip and Intelligent Optoelectronic Systems, Chongqing Municipal Education Commission, Yangtze Normal University,  Chongqing 408100, China\\}

%\author{Sui-Hu Dang}
%\affiliation{School of Electronic Information Engineering, Yangtze Normal University, Chongqing 408100, China\\}
%\affiliation{Key Laboratory of Optical Chip and Intelligent Optoelectronic Systems, Chongqing Municipal Education Commission, Yangtze Normal University,  Chongqing 408100, China\\}

\author{Zhi-Cheng Shi}
\affiliation{Fujian Key Laboratory of Quantum Information and Quantum Optics, Fuzhou University, Fuzhou 350116, China\\}
\affiliation{Department of Physics, Fuzhou University, Fuzhou, 350116, China\\}

\author{Yan Xia}
\email{xia-208@163.com}
\affiliation{Fujian Key Laboratory of Quantum Information and Quantum Optics, Fuzhou University, Fuzhou 350116, China\\}
\affiliation{Department of Physics, Fuzhou University, Fuzhou, 350116, China\\}

\author{Avraham Gover}
\email{gover@eng.tau.ac.il}
\affiliation{Department of Electrical Engineering Physical Electronics,Tel Aviv University, Ramat Aviv 69978, Israel\\}

\begin{abstract}
We propose a protocol that transfers entanglement from an entangled atomic two-level-system (TLS) resource to a pair of free electrons in an energy-sideband ladder via local electron--TLS interactions.
In a controlled rotating-wave regime, closed-form reduced states are derived.
TLS heralding then prepares a maximally entangled electron state in a two-dimensional single-excitation manifold, with a simple dependence on the initial TLS resource entanglement.
Numerical integration of the full bilinear Hamiltonian quantifies the impacts of detuning and pulse shaping and identifies the leading beyond-rotating-wave corrections.
The results establish a heralded route to entangled free electrons and will facilitate further advances in quantum electron optics.
\end{abstract}

\maketitle

Quantum entanglement, the nonlocal correlation in multipartite quantum systems, enables practical applications such as quantum teleportation \cite{luo2019quantum}, quantum cryptography \cite{ekert1991quantum}, quantum communication \cite{bennett1993teleporting}, and quantum computing \cite{gottesman1999demonstrating}, and it supports a wide range of fundamental tests of quantum mechanics \cite{aspect1982experimental,greenberger1990bell,paneru2021experimental}.
Over the past decades, entanglement has been generated across diverse physical platforms, including photons \cite{thomas2022efficient,chen2021quantum,forbes2025heralded}, trapped ions \cite{song2019generation}, neutral atoms \cite{dudarev2003entanglement,wilk2010entanglement,shao2017ground,yang2022sequential,jo2020rydberg}, superconducting circuits \cite{chen2025hardware,campagne2018deterministic,riste2013deterministic,zhang2023generating}, solid-state spins \cite{bernien2013heralded,delteil2016generation}, and cavity quantum electrodynamics \cite{zheng2000efficient,zheng2001one,zheng2002quantum}.
Progress in entanglement generation has been driven by a broad set of mechanisms, including the quantum Zeno effect \cite{wang2008quantum,barontini2015deterministic,nodurft2022generation,nodurft2022generation,liang2015adiabatic}, measurement-feedback control \cite{miki2023generating,francesco2024steady,hashim2025efficient,yamamoto2007feedback,mancini2006markovian,wang2005dynamical}, dissipation engineering \cite{kastoryano2011dissipative,zhang2023generation,su2014scheme,cole2022resource}, and adiabatic passage \cite{chang2020remote,xu2024efficient,carrasco2024dicke}.
Such advances allow the preparation of entangled states spanning Bell states \cite{roos2004bell,yuan2020steady,zou2022bell,weng2025high}, multipartite GHZ states \cite{yang2018deterministic,zhang2024fast}, W states \cite{zou2002generation,zang2016deterministic,li2015generation,peng2021one,deng2006generation}, Dicke states \cite{toyoda2011generation,wu2017generation,noguchi2012generation,thiel2007generation}, and cluster states \cite{zheng2006generation,dong2006generation}, as well as entanglement in continuous-variable and hybrid settings \cite{shinjo2019pulse,meesala2024quantum,knaut2024entanglement,chang2025hybrid}.

More recently, quantum electron optics (QEOs) has emerged as a platform for coherent preparation, manipulation, and measurement of free-electron wave packets via photon-induced near-field electron microscopy (PINEM), electron energy loss spectroscopy (EELS), and cathodoluminescence \cite{reinhardt2020theory,madan2022ultrafast,vanacore2018attosecond,chirita2022transverse,wong2021control,volkov2022spatiotemporal,gorlach2024double,yannai2023lossless,vanacore2020spatio,gorlach2024double,garcia2023spatiotemporal,paschen2023ultrafast,tsesses2023tunable,ruimy2023superradiant}.
QEOs enable ultrafast, nanoscale light--matter spectroscopy and quantum-state engineering by using shaped electron--matter interactions to coherently imprint, probe, and herald quantum correlations in emitted or scattered bosonic fields \cite{di2019probing,kfir2021optical,abad2024electron,sun2023generating,ben2021shaping,karnieli2023quantum,bucher2023free,tsarev2021measurement,garcia2021optical,shiloh2022quantum,huang2023quantum,karnieli2023quantum,arend2025electrons,karnieli2024universal,yang2025unifying,velasco2024radiative}.
The free-electron wave-packet size controls the coupling to an atomic TLS via the free-electron-bound-electron resonant interaction (FEBERI), where the coupling is governed by the QEW longitudinal density modulation \cite{gover2020free}.
Building on this framework, engineered QEWs enable efficient excitation and quantum-state interrogation of TLS \cite{zhang2021quantum,zhang2022quantum,ran2022coherent}, and have motivated a flurry of recent work on related topics \cite{crispin2025probing,zhao2021quantum,yalunin2021tailored,morimoto2021coherent,ruimy2021toward,ratzel2021controlling,zhang2025spontaneous,abad2025quantum}.
More broadly, free-electron interactions can generate entanglement between electron and material excitations such as confined cavity photons and polaritonic or vibrational modes \cite{kfir2019entanglements,yanagimoto2023time,baranes2022free,konevcna2022entangling,mechel2021quantum,rasmussen2024generation,arque2022atomic,baranes2023free,feist2022cavity}.

Despite rapid progress in coherent control and electron--matter interactions, a clear route to preparing entanglement between free electrons remains underexplored.
Free-electron entanglement is of particular interest because it can support super- and subradiant light emission from entangled electron pairs \cite{karnieli2021superradiance}, enable enhanced deterministic photon generation using symmetric multi-electron entangled states \cite{karnieli2023jaynes}, and may enrich electron--matter probes, thereby opening new opportunities for QEOs \cite{ruimy2025free}.
Motivated by this gap, this Letter presents a heralded protocol that transfers entanglement from an atomic TLS pair to two free electrons and provides analytic relations and full-model numerical benchmarks for the generation of electron Bell states.

\textit{System model:}
We consider two free electrons, labeled $A$ and $B$, propagating along spatially separated trajectories and interacting locally with two spatially separated atomic TLSs, with $A$ coupled only to TLS~1 and $B$ only to TLS~2.
Each free electron is described in the standard discrete \textit{energy-sideband} basis used in photon-induced near-field electron microscopy and related phase-modulation settings \cite{ben2021shaping,baranes2022free}.
The electron wave packet is decomposed into components whose energies differ by integer multiples of a reference quantum $\hbar\omega$.
We denote by $\{|n\rangle\}_{n\in\mathbb{Z}}$ the electron energy-sideband basis, where $n$ labels the energy shift $E_0+n\hbar\omega$ with respect to the ``zero-loss" energy $E_0$.
The corresponding sideband-number operator is $\hat{n}=\sum\limits_{n=-\infty}^{\infty} n\,|n\rangle\langle n|$.
Following the standard scattering-matrix description of phase-modulated free electrons, we introduce the commuting electron-energy ladder operator
$\hat{b}=\sum\limits_{n=-\infty}^{\infty}|n-1\rangle\langle n|$, which satisfies the defining actions $\hat{b}|n\rangle=|n-1\rangle$, $\hat{b}^{\dagger}\hat{b}=\hat{b}\hat{b}^{\dagger}=\hat{\mathbb{I}}$, and $[\hat{b},\hat{b}^{\dagger}]=0$.
The free Hamiltonian for the sideband degree of freedom is $\hat{H}_{\mathrm{e}}=\hbar\omega\,\hat{n}$, i.e., adjacent sidebands are separated by $\hbar\omega$ ($E_0$ contributes only a global phase and is omitted).
Each TLS $j\in\{1,2\}$ has ground state $|g\rangle_{j}$, excited state $|e\rangle_{j}$, and transition frequency $\omega_{0}$.
We define the Pauli operators $\hat{\sigma}_{j+}=|e\rangle_{j}\langle g|$, $\hat{\sigma}_{jz}=|e\rangle_{j}\langle e|-|g\rangle_{j}\langle g|$.
The bare TLS Hamiltonian is $\hat{H}_{\mathrm{TLS}}=\sum_{j=1}^{2}\frac{\hbar\omega_{0}}{2}\,\hat{\sigma}_{jz}$
(Algebraic properties of these operators are summarized in Supplementary Sec.~S1. \cite{SM})

When a free electron passes in the near field of a TLS (bound electron), the interaction is Coulombic and can be expressed as a dipole coupling between the TLS dipole operator $\hat{\mathbf{d}}_{j}$ and the electric field generated by the passing (bunched) electron wavefunction at the TLS position, $\mathbf{E}_{\mathrm{e}}(\mathbf{r}_{j},t)$ \cite{ran2022coherent,abad2025quantum}.
At the level of an effective two-level description, one may write
$\hat{H}_{\mathrm{int}}^{(j)}(t)=-\hat{\mathbf{d}}_{j}\cdot \mathbf{E}_{\mathrm{e}}(\mathbf{r}_{j},t)$,
and, after projecting $\hat{\mathbf{d}}_{j}$ onto the TLS subspace and expressing the electron energy exchange in the sideband basis, the interaction Hamiltonian reads
\begin{equation}
\hat{H}_{\mathrm{int}}^{(j)}(t)
=
\hbar\,\mathcal{G}_{j}(t)\,
\bigl(\hat{b}_{\alpha(j)}+\hat{b}_{\alpha(j)}^{\dagger}\bigr)
\bigl(\hat{\sigma}_{j+}+\hat{\sigma}_{j-}\bigr),
\label{eq:sys_Hint_bilinear}
\end{equation}
where $\alpha(j)$ labels the electron addressed by TLS $j$, and $\mathcal{G}_{j}(t)$ is a real-valued effective coupling envelope.
For an electron traversing the near-field region of TLS $j$, the interaction strength is generally position dependent along the trajectory.
A convenient description is therefore to introduce an effective time envelope by mapping the spatial profile onto time through the electron motion.
Assuming an approximately constant longitudinal velocity $v$, one writes the coupling as $\mathcal{G}_{j}(t)\equiv \mathcal{G}_{j}\!\big(z=vt\big)$, so that all spatial dependence enters as a time-dependent modulation during the finite transit.
For a pulse-like interaction of duration $T$, it is convenient to define the integrated coupling (pulse area) $g_{j}(t)\equiv\int_{0}^{t}\!dt'\,\mathcal{G}_{j}(t')$, then $g_{j}\equiv g_{j}(T)$.
This dimensionless quantity is determined by the field experienced along the electron trajectory and provides an operational measure of the interaction strength.
In the present work it controls coherent energy exchange between the electron sideband degree of freedom and the TLS transition.
For simplicity, we adopt an equal-area simplification in which the detailed envelope is replaced by a constant-amplitude pulse with the same $g_{j}$.
Specifically, we take $\mathcal{G}_{j}(t)=G_{0,j}$ for $0\le t\le T$ and $\mathcal{G}_{j}(t)=0$ otherwise, which gives $g_{j}(t)=G_{0,j}\,t$ and $g_{j}=G_{0,j}T$.
This square-pulse model should be understood as a minimal analytic surrogate that preserves the experimentally controlled quantity $g_{j}$, while the full simulations may use the original time envelope $\mathcal{G}_{j}(t)$ when needed.

The total Hamiltonian of the considered system is
\begin{equation}
\hat{H}(t)
=
\hat{H}_{\mathrm{e},A}+\hat{H}_{\mathrm{e},B}
+
\hat{H}_{\mathrm{TLS}}
+
\hat{H}_{\mathrm{int}}^{(1)}(t)
+
\hat{H}_{\mathrm{int}}^{(2)}(t),
\label{eq:sys_total_H}
\end{equation}
with $\hat{H}_{\mathrm{e},A}=\hbar\omega\,\hat{n}_{A}$ and $\hat{H}_{\mathrm{e},B}=\hbar\omega\,\hat{n}_{B}$ acting on the respective electron sideband spaces.
A key control parameter is the detuning between the sideband spacing and the TLS transition, $\Delta \equiv \omega-\omega_{0}$ and $\Omega_{+}\equiv \omega+\omega_{0}$, which naturally separates near-resonant exchange processes (set by $\Delta$) from counter-rotating processes (set by the large frequency $\Omega_{+}$).
In the regime $|\Delta|\ll \omega,\omega_{0}$ and sufficiently weak coupling, Eq.~(\ref{eq:sys_Hint_bilinear}) reduces to the familiar near-resonant exchange form under a controlled rotating-wave approximation (RWA).
The FULL harmonic decomposition and controlled-RWA validity diagnostics are summarized in Supplementary Sec.S2.

On exact resonance, the near-resonant exchange interaction couples only two joint configurations at a time.
The term $\hat{b}\hat{\sigma}_{+}$ maps $|n,g\rangle$ to $|n-1,e\rangle$, while $\hat{b}^{\dagger}\hat{\sigma}_{-}$ maps $|n-1,e\rangle$ back to $|n,g\rangle$.
Therefore, for each integer $n$, the subspace $\mathrm{span}\{|n,g\rangle,\ |n-1,e\rangle\}$ is invariant.
Equivalently, $\mathrm{span}\{|n,e\rangle,\ |n+1,g\rangle\}$ is invariant and describes coherent conversion of a TLS excitation into an electron energy gain.
With the accumulated pulse area $g_j(t)$, the resonant rotating-wave propagator up to time $t$ reads
\begin{equation}
\hat{U}^{\alpha(j)}_{\mathrm{RWA}}(t)
=
\exp\!\left[
-i\,g_j(t)\Big(\hat{b}\hat{\sigma}_{+}+\hat{b}^{\dagger}\hat{\sigma}_{-}\Big)
\right].
\label{eq:A1_U_RWA_time}
\end{equation}
Its action on the joint electron--TLS basis is, for all integers $n$,
\begin{eqnarray}
|n,g\rangle
&\longrightarrow&
\cos\!\big(g(t)\big)\,|n,g\rangle
-i\sin\!\big(g(t)\big)\,|n-1,e\rangle,
\nonumber\\
|n,e\rangle
&\longrightarrow&
\cos\!\big(g(t)\big)\,|n,e\rangle
-i\sin\!\big(g(t)\big)\,|n+1,g\rangle.
\label{eq:A1_map_ne_time}
\end{eqnarray}
Equations~(\ref{eq:A1_map_ne_time}) give a closed-form, time-resolved solution for the resonant rotating-wave dynamics.

We now consider the case where the two electron--TLS interactions are applied over the same time window $0\le t\le T$.
On exact resonance and within the RWA, the interaction-picture Hamiltonian is the sum of two local exchange terms,
\begin{eqnarray}
\hat{H}_{I,\mathrm{tot}}^{\mathrm{(RWA)}}(t)
&=&
\hbar\,\mathcal{G}(t)\,
\left(
\hat{b}_{A}\hat{\sigma}_{1+}+\hat{b}_{B}\hat{\sigma}_{2+}
+ H.c.
\right).
\label{eq:A1_Htot_RWA_time}
\end{eqnarray}
The two local contributions act on disjoint tensor factors and therefore commute at all times, $\left[
\hat{b}_{A}\hat{\sigma}_{1+}+\hat{b}_{A}^{\dagger}\hat{\sigma}_{1-},
\,
\hat{b}_{B}\hat{\sigma}_{2+}+\hat{b}_{B}^{\dagger}\hat{\sigma}_{2-}
\right]=0$.
As a result, the time-ordered propagator factorizes exactly, $\hat{U}_{\mathrm{tot}}^{\mathrm{(RWA)}}(t)=\hat{U}_{A1}^{\mathrm{(RWA)}}(t)\otimes \hat{U}_{B2}^{\mathrm{(RWA)}}(t)$, where each $\hat{U}_{A1}^{\mathrm{(RWA)}}(t)$ and $\hat{U}_{B2}^{\mathrm{(RWA)}}(t)$ is the single-pair propagator introduced in Eq. (\ref{eq:A1_U_RWA_time}).
This exact factorization lets us construct the full four-partite state by applying the single-pair map independently on the two arms.

\paragraph{Heralded entanglement transfer: }
We take both electrons initially in the same energy-sideband state $|\psi_{E}(0)\rangle=|n_{0}\rangle_{A}\,|n_{0}\rangle_{B}$, and we prepare the TLS pair in the single-excitation Bell state,
\begin{eqnarray}
|\Psi^{+}\rangle_{12}
&=&
\frac{1}{\sqrt{2}}\left(|e g\rangle_{12}+|g e\rangle_{12}\right),
\label{eq:A1_TLS_Bell_init_time}
\end{eqnarray}
Then the initial electron--TLS state is $|\Psi(0)\rangle=|\psi_{E}(0)\rangle\otimes|\Psi^{+}\rangle_{12}$.
The full time-dependent four-partite state within the pulse follows by applying the local maps in Eq.~(\ref{eq:A1_map_ne_time}) to each component of $|\Psi^{+}\rangle_{12}$.
The evolution of electron--TLS state is then given by
\begin{eqnarray}
|\Psi(t)\rangle
&=&
\frac{1}{\sqrt{2}}
\Big[
|\Phi_{eg}(t)\rangle
+
|\Phi_{ge}(t)\rangle
\Big],
\label{eq:A1_Psi_t_exact_def}
\end{eqnarray}
where
\begin{eqnarray}
&|\Phi_{eg}(t)\rangle
&=
\left[c(t)\,|n_{0}\rangle_{A}|e\rangle_{1}-i s(t)\,|n_{0}\!+\!1\rangle_{A}|g\rangle_{1}\right]
\nonumber\\
&&
\otimes
\left[c(t)\,|n_{0}\rangle_{B}|g\rangle_{2}-i s(t)\,|n_{0}\!-\!1\rangle_{B}|e\rangle_{2}\right].
\label{eq:A1_evolve_eg_component_time}
\end{eqnarray}
and
\begin{eqnarray}
&|\Phi_{ge}(t)\rangle
&=
\left[c(t)\,|n_{0}\rangle_{A}|g\rangle_{1}-i s(t)\,|n_{0}\!-\!1\rangle_{A}|e\rangle_{1}\right]
\nonumber\\
&&
\otimes
\left[c(t)\,|n_{0}\rangle_{B}|e\rangle_{2}-i s(t)\,|n_{0}\!+\!1\rangle_{B}|g\rangle_{2}\right].
\label{eq:A1_evolve_ge_component_time}
\end{eqnarray}
with $c(t)\equiv\cos\!\big(g(t)\big)$ and $s(t)\equiv\sin\!\big(g(t)\big)$.
It is convenient to introduce the shorthand electron basis states
$|00\rangle \equiv  |n_0\rangle_A|n_0\rangle_B, |+-\rangle \equiv  |n_0\!+\!1\rangle_A|n_0\!-\!1\rangle_B,
|\pm0\rangle \equiv  |n_0\!\pm\!1\rangle_A|n_0\rangle_B,  |0\pm\rangle \equiv  |n_0\rangle_A|n_0\pm 1\rangle_B$.

We track how the initial TLS entanglement is redistributed during the interaction.
The unconditional TLS reduced state is $\hat{\rho}_{12}(t)=\mathrm{Tr}_{AB}\!\left[|\Psi(t)\rangle\langle\Psi(t)|\right]$.
Orthogonality between distinct electron sideband configurations removes many cross terms, and $\hat{\rho}_{12}(t)$ takes an $X$-state form in the basis $\{|gg\rangle,\ |ge\rangle,\ |eg\rangle,\ |ee\rangle\}$.
The nonzero matrix elements are
\begin{eqnarray}
\rho_{gg,gg}(t)
&=&
s^{2}(t)c^{2}(t),
\qquad
\rho_{ee,ee}(t)
=
s^{2}(t)c^{2}(t),
\label{eq:A1_TLS_pop_gg_ee_time}
\nonumber\\
\rho_{eg,eg}(t)
&=&
\frac{c^{4}(t)+s^{4}(t)}{2},
\rho_{ge,ge}(t)
=
\frac{c^{4}(t)+s^{4}(t)}{2},
\label{eq:A1_TLS_pop_eg_ge_time}
\nonumber\\
\rho_{eg,ge}(t)
&=&
\rho_{ge,eg}(t)
=
\frac{c^{4}(t)}{2}.
\label{eq:A1_TLS_coherence_time}
\end{eqnarray}
For an $X$ state, the TLS entanglement is quantified by the concurrence
\begin{eqnarray}
\mathcal{C}_{12}(t)
&=&
2\max\{
0,
|\rho_{eg,ge}(t)|-\sqrt{\rho_{gg,gg}(t)\rho_{ee,ee}(t)}
\}.\ \ \
\label{eq:A1_TLS_concurrence_Xformula_time}
\end{eqnarray}
Substituting Eq.~(\ref{eq:A1_TLS_pop_gg_ee_time}) yields
\begin{eqnarray}
\mathcal{C}_{12}(t)
=
\max\!\left\{
0,\,
c^{4}(t)-2s^{2}(t)c^{2}(t)
\right\}.
\label{eq:A1_TLS_concurrence_closed_time}
\end{eqnarray}
A general resource-state reconstruction and closed-form reduced-state expressions for $\rho_{12}(t)$ and $\rho_{AB}(t)$  are derived in Supplementary Sec.S3.

Projecting onto the outcome $|gg\rangle_{12}$ retains only the terms where both TLSs are in the ground state.
The corresponding unnormalized component is $|\Psi_{gg}(t)\rangle=-i\,s(t)c(t)|\Psi^{+}_{E}\rangle\otimes|gg\rangle_{12}$, where the electron Bell state $|\Psi^{+}_{E}\rangle\equiv\frac{|+0\rangle+|0+\rangle}{\sqrt{2}}$.
The heralding probability is therefore
\begin{eqnarray}
P_{gg}(t)
&\equiv&
\langle \Psi(t)|
\left(\mathbb{I}_{AB}\otimes |gg\rangle\langle gg|_{12}\right)
|\Psi(t)\rangle  \nonumber\\
&=&
s^{2}(t)c^{2}(t)
=
\sin^{2}\!\big(g(t)\big)\cos^{2}\!\big(g(t)\big).
\label{eq:A1_Pgg_t_closed_form}
\end{eqnarray}
Under the square-pulse approximation, $g(t)=G_{0}t$ for $0\le t\le T$,
\begin{eqnarray}
P_{gg}(t)
&=&
\sin^{2}\!\left(G_{0}t\right)\cos^{2}\!\left(G_{0}t\right).
\label{eq:A1_Pgg_square_time}
\end{eqnarray}

Using Eqs.~(\ref{eq:A1_evolve_eg_component_time}) and (\ref{eq:A1_evolve_ge_component_time}), the four-partite pure state
$|\Psi(t)\rangle$ can be decomposed in the TLS computational basis as
\begin{eqnarray}
|\Psi(t)\rangle
=
|\Psi^{(gg)}_{AB}(t)\rangle \,|gg\rangle_{12}
+
|\Psi^{(ee)}_{AB}(t)\rangle \,|ee\rangle_{12}  \nonumber\\
+
|\Psi^{(eg)}_{AB}(t)\rangle \,|eg\rangle_{12}
+
|\Psi^{(ge)}_{AB}(t)\rangle \,|ge\rangle_{12},
\label{eq:Psi_decomposition_TLS}
\end{eqnarray}
where the (unnormalized) TLS-conditional electron amplitudes are
\begin{eqnarray}
|\Psi^{(gg)}_{AB}(t)\rangle
&=&
\frac{-i\,s(t)c(t)}{\sqrt{2}}\Big(|+0\rangle+|0+\rangle\Big),
\label{eq:Psi_gg_AB}
\nonumber\\
|\Psi^{(ee)}_{AB}(t)\rangle
&=&
\frac{-i\,s(t)c(t)}{\sqrt{2}}\Big(|-0\rangle+|0-\rangle\Big),
\label{eq:Psi_ee_AB}
\nonumber\\
|\Psi^{(eg)}_{AB}(t)\rangle
&=&
\frac{1}{\sqrt{2}}\Big(c^{2}(t)\,|00\rangle-s^{2}(t)\,|-+\rangle\Big),
\label{eq:Psi_eg_AB}
\nonumber\\
|\Psi^{(ge)}_{AB}(t)\rangle
&=&
\frac{1}{\sqrt{2}}\Big(c^{2}(t)\,|00\rangle-s^{2}(t)\,|+-\rangle\Big).
\label{eq:Psi_ge_AB}
\end{eqnarray}

Since the TLS basis states are orthonormal, tracing out the TLSs yields the unconditional electron reduced state $\hat{\rho}_{AB}(t)\equiv\mathrm{Tr}_{12}\!\left[\,|\Psi(t)\rangle\langle\Psi(t)|\,\right]$.
A standard computable entanglement monotone is the negativity, $\mathcal{N}_{AB}(t)
\equiv
\frac{\left\|\hat{\rho}_{AB}^{\,T_B}(t)\right\|_{1}-1}{2}$,
where $T_B$ denotes partial transpose with respect to electron $B$ and $\|\cdot\|_1$ is the trace norm.
The corresponding logarithmic negativity is
\begin{equation}
E^{AB}_{\mathcal{N}}(t)
\equiv
\log_{2}\!\left\|\hat{\rho}_{AB}^{\,T_B}(t)\right\|_{1}.
\label{eq:logneg_def}
\end{equation}

\begin{figure}[t]
\centering
\includegraphics[width=0.45\textwidth]{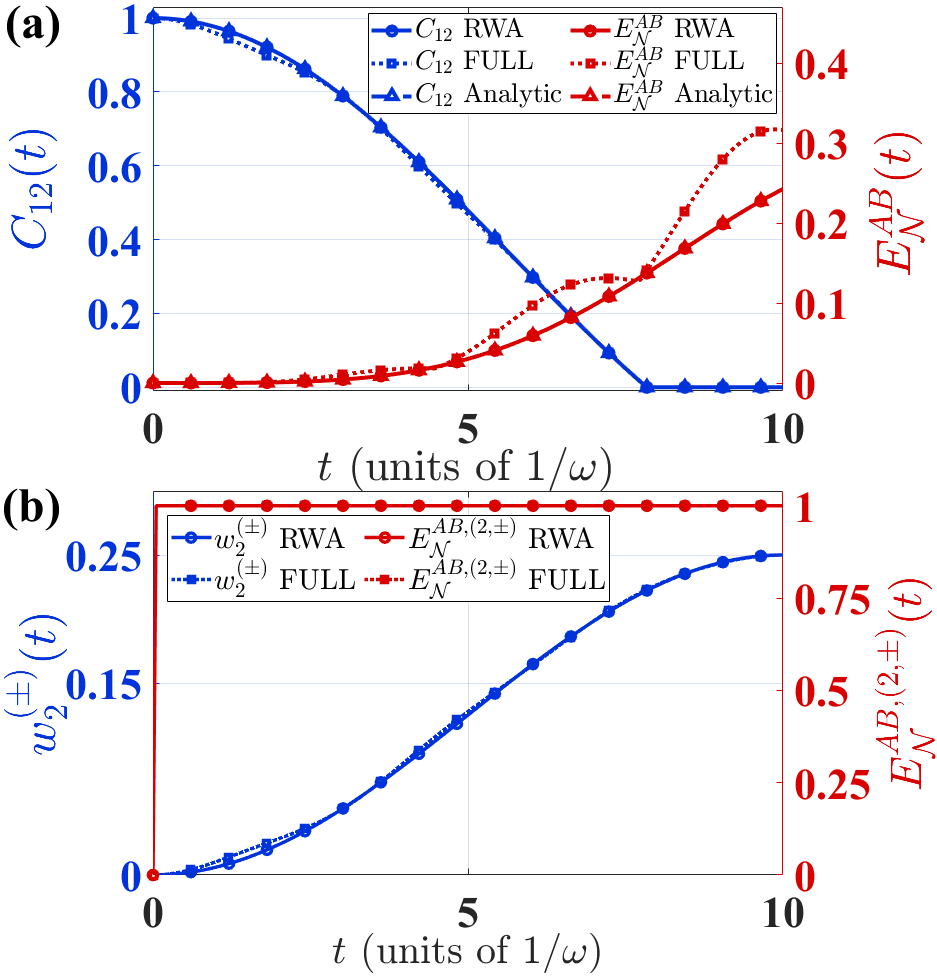}
\caption{
Entanglement-transfer dynamics between the TLSs and the free-electrons.
(a) Unconditional entanglement measures.
Left axis: TLS concurrence $\mathcal{C}_{12}(t)$.
Right axis: electron logarithmic negativity $E_{\mathcal{N}}^{AB}(t)$, shown for the RWA numerics, the full bilinear model (FULL) numerics, and the resonant closed-form RWA benchmark.
(b) Heralded extraction into the two-dimensional target manifolds $\mathcal{H}_{2}^{(\pm)}$.
Left axis: unconditional target-manifold weight $w_{2}^{(\pm)}(t)$.
Right axis: logarithmic negativity $E_{\mathcal{N}}^{AB,(2,\pm)}(t)$ of the normalized projected state $\hat{\rho}_{AB}^{(2,\pm)}(t)$.
Parameters: $\Delta=0$ and pulse area $g=\pi/4$.
}
\label{fig:ent_transfer_dynamics}
\end{figure}

The protocol targets the two-dimensional electron manifold
$\mathcal{H}^{(\pm)}_{2}\equiv\mathrm{span}\left\{|\pm0\rangle,\ |0\pm\rangle\right\},
\hat{P}^{(\pm)}_{2}\equiv |\pm0\rangle\langle \pm0|+|0\pm\rangle\langle0\pm|$.
From Eq.~(\ref{eq:Psi_gg_AB}) and the orthogonality of distinct sideband configurations, only the $|gg\rangle_{12}$ (for $+$) and  $|ee\rangle_{12}$ (for $-$) branch have support in $\mathcal{H}_{2}^{(\pm)}$, hence
\begin{equation}
w_{2}^{(\pm)}(t)
\equiv
\mathrm{Tr}\!\left[\hat{P}_{2}^{(\pm)}\hat{\rho}_{AB}(t)\right]
=
s^{2}(t)c^{2}(t)=P_{gg}(t).
\label{eq:w2plus_of_t}
\end{equation}
Moreover, the normalized state within $\mathcal{H}_{2}^{(\pm)}$ is
\begin{eqnarray}
\hat{\rho}^{(2,\pm)}_{AB}(t)
\equiv
\frac{\hat{P}_{2}^{(\pm)}\hat{\rho}_{AB}(t)\hat{P}_{2}^{(\pm)}}{w_{2}^{(\pm)}(t)}
=
|\Psi^{\pm}_{E}\rangle\langle\Psi^{\pm}_{E}|,
\label{eq:rhoAB_restricted_plus}
\end{eqnarray}
where $|\Psi^{-}_{E}\rangle\equiv\frac{|-0\rangle+|0-\rangle}{\sqrt{2}}$, which is independent of $t$ after renormalization, while the operational weight is $w_{2}^{(\pm)}(t)$.
Therefore, the same interaction transfers the initial TLS Bell entanglement into two symmetric electron Bell states residing in the $\pm 1$ sideband manifolds, each with the same time-dependent weight $s^{2}(t)c^{2}(t)$ in the unconditional electron state.

Therefore, the entanglement contained in the normalized restricted state
$\hat{\rho}^{(2,\pm)}_{AB}(t)$ can be quantified using the same logarithmic negativity as in
Eq.~(\ref{eq:logneg_def}), now applied within the two-dimensional identification
$\mathcal{H}^{(\pm)}_{2}$.
We define
\begin{eqnarray}
E^{AB,(2,\pm)}_{\mathcal{N}}(t)
\equiv
\log_{2}\!\left\|
\left[\hat{\rho}^{(2,\pm)}_{AB}(t)\right]^{T_{B}}
\right\|_{1}.
\label{eq:logneg_restricted_def}
\end{eqnarray}
Using Eq.~(\ref{eq:rhoAB_restricted_plus}), $\hat{\rho}^{(2,\pm)}_{AB}(t)=|\Psi^{\pm}_{E}\rangle\langle\Psi^{\pm}_{E}|$ is a Bell state for all $t$ after renormalization.
Hence the restricted logarithmic negativity is time independent,
\begin{equation}
E^{AB,(2,\pm)}_{\mathcal{N}}(t)
=
\log_{2}\!\left\|
\left(|\Psi^{\pm}_{E}\rangle\langle\Psi^{\pm}_{E}|\right)^{T_{B}}
\right\|_{1}
=
1,
\label{eq:logneg_restricted_equals_1}
\end{equation}
equivalently implying unit concurrence under the same two-qubit identification.
While the normalized restricted entanglement is constant, its  operational contribution to the
unconditional electron state is weighted by the target-manifold population
$w_{2}^{(\pm)}(t)$ in Eq.~(\ref{eq:w2plus_of_t}),
which carries the full time dependence through $g(t)$.

We now consider a  {general} entangled TLS resource in the single-excitation manifold, so as to quantify how the {initial TLS entanglement} controls the amount of entanglement that can be transferred to the free-electron pair.
Specifically, we prepare the initial atomic state in
\begin{eqnarray}
\ket{\psi_{12}(0)}
=
\alpha\ket{eg}
+
\beta e^{i\phi}\ket{ge}, \
|\alpha|^{2}+|\beta|^{2}=1,
\label{eq:main_TLS_resource_alpha_beta}
\end{eqnarray}
whose entanglement is quantified by the concurrence $C_{12}(0)=2|\alpha\beta|$.
Under the symmetric RWA exchange, the evolved state admits a branch decomposition over TLS outcomes.
Conditioned on the relevant postselection branch (TLS in $\ket{gg}$ or $\ket{ee}$), the reduced electron state has a dominant contribution in a two-dimensional single-excitation subspace, $\mathcal{H}_{2}^{(\pm)}$.
Projecting onto this subspace and renormalizing yields an effective $2\times 2$ density matrix of the form
\begin{eqnarray}
\rho_{AB}^{(2,\pm)}(T)
=
\left(
\begin{array}{cc}
|\alpha|^{2} & \alpha\beta^{*} e^{-i\phi} \\
\alpha^{*}\beta e^{i\phi} & |\beta|^{2}
\end{array}
\right),
\label{eq:main_rhoAB_2pm_form}
\end{eqnarray}
up to an overall success weight that cancels upon renormalization.
For any state supported on $\mathcal{H}_{2}^{(\pm)}$, the partial transpose has trace norm
$\big\|(\rho_{AB}^{(2,\pm)})^{T_{B}}\big\|_{1}=1+2|\rho_{12}|$, hence the log-negativity is
\begin{eqnarray}
E_{\mathcal N}^{AB,(2,\pm)}(T)
&=&
\log_{2}\!\big(1+C_{12}(0)\big) \nonumber\\
&=&
\log_2\!\Big(1+2\alpha\sqrt{1-\alpha^2}\Big),
\label{eq:EN2pm_C12_closed}
\end{eqnarray}
which is independent of the relative phase $\phi$.
The corresponding general-resource reconstruction and reduced-state diagnostics used to evaluate both unconditional and heralded electron entanglement are provided in Supplementary S3.

\paragraph{Numerical results and validation.}
To benchmark the time-resolved entanglement-transfer dynamics, we solve the interaction-picture Schr\"odinger equation $i\hbar\frac{d}{dt}|\Psi(t)\rangle=\hat{H}_{I}(t)\,|\Psi(t)\rangle$.
Figure~\ref{fig:ent_transfer_dynamics} shows the time-resolved entanglement-transfer dynamics under both the RWA generator in Eq.~(\ref{eq:A1_Htot_RWA_time}) and the full bilinear interaction.
Panel~(a) shows that the numerical curves reproduce the resonant closed-form benchmarks for the TLS concurrence and the electron logarithmic negativity.
In particular, the RWA numerics (solid) coincide with the analytic RWA prediction (triangles) for
$\mathcal{C}_{12}(t)$ in Eq.~(\ref{eq:A1_TLS_concurrence_closed_time}) and for the electron entanglement monotone
$E^{AB}_{\mathcal{N}}(t)$ computed from $\hat{\rho}_{AB}(t)$ via Eq. (\ref{eq:logneg_def}),
while the FULL numerics (dotted) track the same trend and quantify the controlled beyond-RWA corrections.
Panel~(b) confirms the heralded picture.
The target-manifold population follows the analytic weight
$w_{2}^{(\pm)}(t)$ in Eq.~(\ref{eq:w2plus_of_t}),
and the renormalized projected state $\hat{\rho}^{(2,\pm)}_{AB}(t)$ remains a Bell state, so that
$E^{AB,(2,\pm)}_{\mathcal{N}}(t)=1$ in Eq.~(\ref{eq:logneg_restricted_equals_1}) throughout the interaction.
These results verify that the protocol generates heralded free-electron Bell entanglement, while providing a quantitative benchmark for the accuracy of the RWA relative to the full bilinear model.

A notable feature in Fig.~\ref{fig:ent_transfer_dynamics}(a) is that the TLS concurrence $\mathcal{C}_{12}(t)$ reaches zero at a finite time (around $t\simeq 7.8/\omega$), whereas the electron entanglement continues to increase afterwards.
This behavior does not contradict entanglement transfer.
Here $\mathcal{C}_{12}(t)$ quantifies internal entanglement within the reduced two-TLS state $\hat{\rho}_{12}(t)$, so $\mathcal{C}_{12}(t)=0$ only indicates that $\hat{\rho}_{12}(t)$ has become separable.
The resonant RWA analytics predict a finite-time ``sudden death'' of $\mathcal{C}_{12}(t)$ for the chosen pulse area, and the predicted disappearance time agrees with the numerical trace in panel~(a) for $g=\pi/4$ and $T=10$ (giving $t=\frac{1}{G_{0}}\arcsin\!\left(\frac{1}{\sqrt{3}}\right)\simeq 7.8/\omega$).
Physically, the TLS pair can lose its internal concurrence because it becomes correlated with the electron degrees of freedom, which increases the mixedness of $\hat{\rho}_{12}(t)$ and can drive the two-qubit concurrence to zero.
Meanwhile, the interaction continues to populate the entangled $\pm1$ sideband components of the electron state, so the electron pair can still gain bipartite entanglement even after $\mathcal{C}_{12}(t)$ vanishes.
This is consistent with Fig.~\ref{fig:ent_transfer_dynamics}(b), where the weight transferred into the two-dimensional entangled manifolds keeps growing beyond the concurrence-disappearance time.

\begin{figure}[t]
\centering
\includegraphics[width=0.46\textwidth]{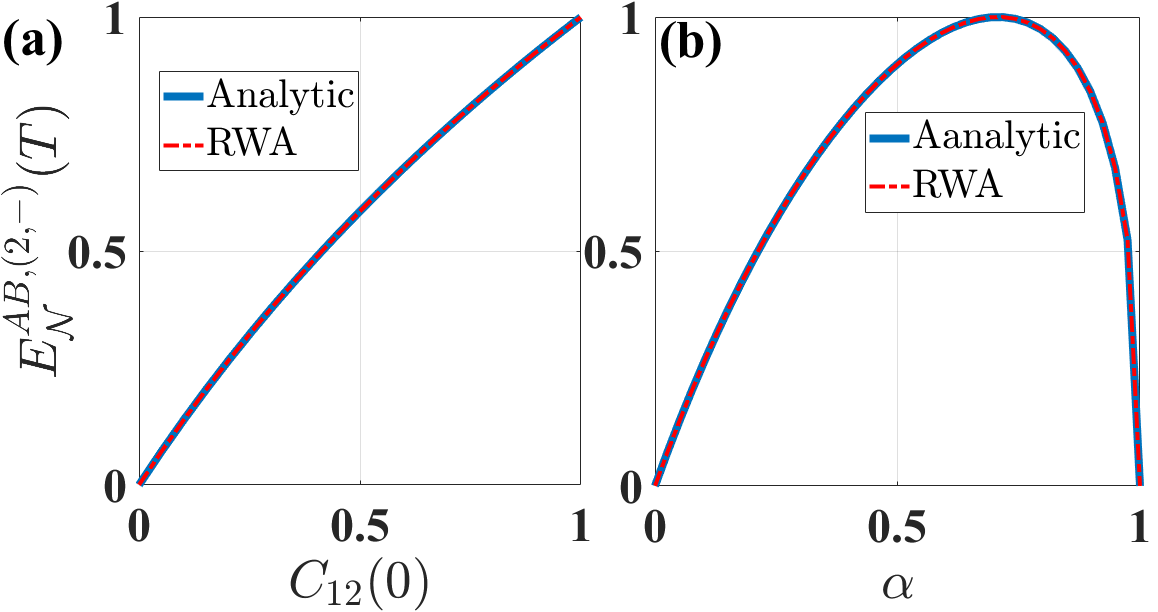}
\caption{
Closed-form relation between the postselected electron log-negativity and the initial TLS concurrence.
(a) shows $E_{\mathcal N}^{AB,(2,-)}(T)$ versus $C_{12}(0)$, where the analytic prediction in Eq. (\ref{eq:EN2pm_C12_closed}) (solid) is compared with numerical RWA evolution (dashed). In the symmetric resonant case the $(2,+)$ and $(2,-)$ branches coincide.
(b) displays the same quantity as a function of the TLS amplitude $\alpha$ for the initial resource.
}
\label{fig:EN2pm_closed}
\end{figure}

Figure~\ref{fig:EN2pm_closed} provides a direct numerical validation of the entanglement transfer in Eq.~(\ref{eq:EN2pm_C12_closed}).
In panel (a), the postselected free-electron entanglement in the $\mathcal{H}_{2}^{(\pm)}$ channel, $E_{\mathcal N}^{AB,(2,\pm)}(T)$, collapses onto a  {single universal curve} when plotted against the {input} TLS entanglement $C_{12}(0)$: the RWA simulation is indistinguishable from the analytic prediction.
This collapse is the hallmark of entanglement transfer in the controlled symmetric setting.
Conditioned on occupying $\mathcal{H}_{2}^{(\pm)}$, the protocol implements an effective two-qubit mapping from the TLS single-excitation subspace to the electronic single-excitation manifold, so the output entanglement is fixed solely by the entanglement content of the resource state, rather than by its microscopic parametrization.
Panel (b) shows the same final electron entanglement as a function of the microscopic amplitude $\alpha$ in Eq.~(\ref{eq:main_TLS_resource_alpha_beta}).
The apparent nonlinearity is not an additional dynamical effect; it follows entirely from the monotone mapping between the resource parameter and the TLS concurrence,
$C_{12}(0)=2|\alpha\beta|$, together with the symmetry $\alpha\leftrightarrow\beta$.
The analytic--numeric agreement confirms that, in the $\mathcal{H}_{2}^{(\pm)}$ transfer channel, the amount of free-electron entanglement generated at time $T$ is completely determined by the initial TLS entanglement through Eq.~(\ref{eq:EN2pm_C12_closed}), i.e., the protocol transfers entanglement from the TLS resource to the electrons.

We want to emphasise that the transfer efficiency is sensitive to the electron--TLS detuning $\Delta=\omega-\omega_{0}$.
Increasing $|\Delta|$ would suppress the coherent exchange that populates the entangled $\pm1$-sideband manifolds, thereby reducing the final free-electron entanglement.
Beyond the RWA, the dominant systematic correction is captured by a Bloch--Siegert renormalization of the effective detuning, while the remaining deviation is governed by a weak heralded leakage whose fixed-area tail follows a controlled $T^{-2}$ scaling.
Detailed derivations, reduced-state reconstructions, and the controlled beyond-RWA analysis are provided in Supplementary Sec.S4.

\textit{Conclusion and outlook:}
We presented a protocol that transfers entanglement from an entangled atomic pair to a pair of free electrons and certifies it by heralding on the atomic measurement outcomes.
The transfer relies on local exchange interactions that map a single atomic excitation into a shared single-excitation electronic manifold, producing entangled Bell electron states conditioned on successful heralding.
The results provide compact analytical relations between the input atomic resource and the heralded electronic entanglement, and they are validated by numerical simulations of the full interaction model.
Detuning should be carefully controlled because it can reduce the attainable electron entanglement.
Interaction-envelope design provides a practical knob to improve performance at fixed interaction strength.
Beyond the rotating-wave regime, the dominant deviations can be summarized by an effective detuning shift and a weak leakage channel.
Entangled electron Bell states provide a concrete resource for exploring collective radiative effects with free-electron pairs, including super- and subradiant emission signatures, and for extending electron--matter interrogation beyond single-particle probes \cite{karnieli2021superradiance,ruimy2025free}.
A natural extension is to move beyond the discrete-sideband plane-wave model and treat realistic electron wave packets, which set the effective interaction duration and spectral selectivity of the coupling and therefore impact both transfer efficiency and readout via the measured energy spectrum \cite{gover2020free,zhang2021quantum,zhang2022quantum,ran2022coherent}.

\
\noindent {\bf Acknowledgments}
This work is supported by the Natural Science Foundation of Chongqing (Grant No. CSTB2025NSCQ-GPX0416) and the Science and Technology Research Program of Chongqing Municipal Education Commission (Grant No. KJQN202401437).
Y.-D. L. was supported by the Science and Technology Research Program of Chongqing Municipal Education Commission (Grant No. KJZD-K202401403) and the scientific research project of the Science and Technology Bureau of Fuling (Grant No. FLKJ2025AAG2003).
S. L. was supported the Natural Science Foundation of Hubei Province (Grant No. 2024AFB200).
Z.-C.~S. was supported by the National Natural Science Foundation of China (Grant No. 62571129) and the Natural Science Foundation of Fujian Province (Grant No. 2025J01456).
Y.~X. was supported by the National Natural Science Foundation of China (Grant Nos. 11575045 and 11874114), the Natural Science Funds for Distinguished Young Scholar of Fujian Province (Grant No. 2020J06011), and a project from Fuzhou University (Grant No. JG2020001-2).
A.G. and R.I. acknowledge the support of the Israel Science Foundation (Grant No. 2992124).

\bibliographystyle{unsrt}
\bibliography{references}
%\bibliographystyle{apsrev4-1}

%\clearpage
%\begin{widetext}
%
%
%\appendix
%\paragraph{Numerical verification:}
%To benchmark the time-resolved entanglement-transfer dynamics, we numerically propagate the four-partite state vector
%
%\clearpage
%\end{widetext}

\end{document}